\documentclass[journal,twoside]{IEEEtran}

\usepackage{cite}
\usepackage{amsmath,amssymb,amsfonts}
\usepackage{algorithmic}
\usepackage{graphicx}
\usepackage{textcomp}
\usepackage{xcolor}
\usepackage{booktabs}
\usepackage{fancyhdr} 

\usepackage{soul}
\newcommand{\addtext}[1]{\textcolor{black}{#1}}

\usepackage{xcolor}

\fancypagestyle{ieee}{ 
    \fancyhf{} 
    \fancyhead[C]{
    \footnotesize This article has been accepted for publication in IEEE Transactions on Circuits and Systems–I: Regular Papers. This is the author’s version which has not been fully edited and content may change prior to final publication. Citation information: DOI 10.1109/TCSI.2024.3522351}
    \fancyfoot[C]{
    \footnotesize \copyright~2024 IEEE.  Personal use of this material is permitted.  Permission from IEEE must be obtained for all other uses, in any current or future media, including reprinting/republishing this material for advertising or promotional purposes, creating new collective works, for resale or redistribution to servers or lists, or reuse of any copyrighted component of this work in other works.}

}

\def\BibTeX{{\rm B\kern-.05em{\sc i\kern-.025em b}\kern-.08em
    T\kern-.1667em\lower.7ex\hbox{E}\kern-.125emX}}

\begin{document}

\bstctlcite{IEEEexample:BSTcontrol}

\title{TrIM, Triangular Input Movement Systolic Array for Convolutional Neural Networks: \\ Architecture and Hardware Implementation

\thanks{This work was supported by EPSRC FORTE Programme (Grant No. EP/R024642/2) and by the RAEng Chair in Emerging Technologies (Grant No. CiET1819/2/93).}
\thanks{C. Sestito, S. Agwa and T. Prodromakis are with the Centre for 
Electronics Frontiers, Institute for Integrated Micro and Nano Systems,
School of Engineering, The University of Edinburgh, EH9 3BF, Edinburgh, 
United Kingdom. (e-mails: csestito@ed.ac.uk; shady.agwa@ed.ac.uk; 
t.prodromakis@ed.ac.uk).}

}
\author{Cristian~Sestito~\IEEEmembership{Member,~IEEE,} 
        Shady~Agwa~\IEEEmembership{Member,~IEEE,} 
        and~Themis~Prodromakis~\IEEEmembership{Senior Member,~IEEE.}}

\maketitle 
\thispagestyle{ieee} 

\begin{abstract}
Modern hardware architectures for Convolutional Neural Networks (CNNs), other than targeting high performance, aim at dissipating limited energy. Reducing the data movement cost between the computing cores and the memory is a way to mitigate the energy consumption. Systolic arrays are suitable architectures to achieve this objective: they use multiple processing elements that communicate each other to maximize data utilization, based on proper dataflows like the weight stationary and row stationary. Motivated by this, we have proposed TrIM, an innovative dataflow based on a triangular movement of inputs, and capable to reduce the number of memory accesses by one order of magnitude when compared to state-of-the-art systolic arrays.  
In this paper, we present a TrIM-based hardware architecture for CNNs. As a showcase, the accelerator is implemented onto a Field Programmable Gate Array (FPGA) to execute the VGG-16 \addtext{and AlexNet CNNs}. The architecture achieves a peak throughput of 453.6 Giga Operations per Second, outperforming a state-of-the-art row stationary systolic array up to $\sim 3 \times$ in terms of memory accesses, and being up to $ \sim 11.9 \times$ more energy-efficient than other FPGA accelerators.
\end{abstract}


\section{Introduction}
\begin{figure}
\includegraphics[width=\linewidth]{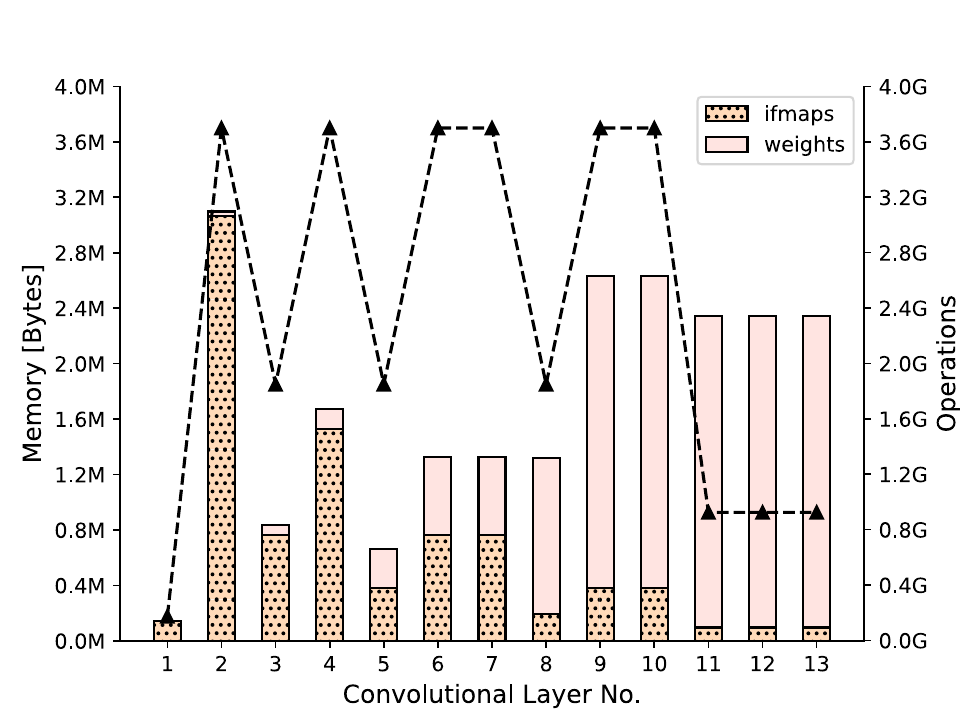}
\centering
\caption{VGG-16 CNN: Memory requirements and operations for each Convolutional Layer (CL). These metrics refer to the inference step at which a generic 224$\times$224 RGB image must be subjected. The memory requirements, expressed in MBytes, are represented by bars. For each bar, both the ifmaps and weights memory are reported. The number of operations, in billions, are highlighted by the dashed line.}
\label{VGG16_Example}
\end{figure}

\IEEEPARstart{C}{onvolutional} Neural Networks (CNNs) are leading AI models in various domains, ranging from computer vision\cite{Spagnolo_23,Emami_21} to speech recognition\cite{Kriman_20,Mustaqeem_20}. Their wide applicability is justified by the use of convolutions, which infer locality between neurons, thus mimicking the behavior of human visual system to identify features from inputs\cite{Zewen_22}. However, this translates in: (i) high memory requirements to manage multi-dimensional feature maps (fmaps); (ii) billions of computations owing to the large fmaps to fulfill high standards of image resolution and accuracy\cite{Sze_17}.
For instance, the well-known VGG-16 CNN\cite{Simonyan_15} requires up to $\sim$22.7 MB of memory to deal with 8-bit input fmaps (ifmaps) and weights, over a total of $\sim$30.7 billions of operations on 224$\times$224 RGB images. Fig.~\ref{VGG16_Example} details the breakdown of memory requirements and computations of the different Convolutional Layers (CLs). Since former CLs process large fmaps, they require massive memory for inputs, as well as computations. Conversely, deeper CLs extract features requiring a dominant contribution of weights.  

In order to follow the fast pace dictated by CNNs' explosion, hardware designers are continuously struggling to guarantee energy-efficient architectures, capable to offer real-time activity as well as low-energy footprint. While the former can be met by means of high-parallelized engines, the latter require to tackle the Von Neumann bottleneck, or \textit{memory wall}\cite{Zou_21}. The major energy-draining source is the communication between the parallel computing cores and the memory, either in the form of on-chip Static Random Access Memory (SRAM) or as an external Dynamic Random Access Memory (DRAM). For example, the relative energy cost for a 32-bit SRAM read is estimated to be 5 pJ for a 45nm process at 0.9V, which in turn dramatically increases to 640 pJ for a 32-bit DRAM read\cite{Horowitz_14}. Overall, the DRAM contribution is $\sim$200$\times$ higher than a 32-bit multiplication performed by the computing core. Among different strategies, on-chip data reuse allows to reduce the number of DRAM reads, by locally storing fmaps and weights to maximize their utilization before being discarded\cite{Lin_18}. 

Systolic Arrays (SAs)\cite{Xu_23,Yuzuguler_23} are promising candidates to mitigate the Von Neumann bottleneck by maximizing data utilization. These architectures consist of an array of Processing Elements (PEs) that mainly perform Multiply-Accumulations (MACs) between inputs and weights. These computing resources are connected each other to ensure two levels of reuse: (i) reuse at the PE level, where one or more types of data (input, weight, partial sum (psum)) are kept \textit{stationary} as long as required; (ii) reuse at the SA level, where one or more types of data are rhythmically moved in specific directions, thus being consumed by different PEs for a certain number of cycles. The reuse at the PE level dictates the specific \textit{dataflow} of the SA: therefore, it is possible to have Input-Stationary (IS)\cite{Wang_22}, Weight-Stationary (WS)\cite{Xu_20}, or Output-Stationary (OS) SAs \cite{Xu_22}, other than mixed combinations of them\cite{Chen_16,Xin_17}. 

While SAs easily manage linear layers, where straightforward MACs are performed between inputs and weights, they are not suitable to directly cope with the workflow exhibited by CLs. Hence, several architectures have converted convolutions into matrix multiplications (Conv-to-GeMM)\cite{Jouppi_17,Fornt_23,Ortega_24,Wu_24,Nandigama_23}. In such an approach, data redundancy is required, which in turn hurts the memory capacity and the main memory accesses.
In another direction, GeMM-free SAs have been proposed\cite{Chen_17,Zhang_24,Sun_23,Liu_19}. For instance, Eyeriss\cite{Chen_17} has introduced a different dataflow named Row Stationary (RS), where inputs and weights are reused in rows at the PE level, by preserving the locality dictated by the convolutional workflow. However, such a dataflow requires many PEs as well as data redundancy at the SA level.

To avoid Conv-to-GeMM and maximize local data utilization, we have proposed a novel dataflow named Triangular Input Movement (TrIM)\cite{Sestito_24} where inputs follows a triangular movement, while weights are kept fixed at the PE level. According to a design space exploration, TrIM has shown one order of magnitude saving in terms of memory accesses when compared to the GeMM-based WS dataflow, while achieving $81.8\%$ throughput improvement when compared to the RS dataflow.

In this work, we propose an AI hardware architecture based on TrIM, dealing with multi-dimensional CLs for CNNs. As a case study, the TrIM architecture is implemented onto a Field Programmable Gate Array (FPGA) to accelerate the VGG-16 \addtext{and AlexNet CNNs} at 150 MHz clock frequency. The design, consisting of 1512 PEs, exhibits a peak throughput of 453.6 GOPs/s, by dissipating $\sim$4.3 W of power. In addition, the TrIM architecture outperforms Eyeriss\cite{Chen_17} by up to $\sim 3 \times$ in terms of memory accesses, and shows the best energy efficiency among state-of-the-art FPGA counterparts.

The main contributions of this work can be summarized as follows:
\begin{itemize}
    \item A high-level scalable TrIM-based AI hardware architecture is presented. \addtext{This includes innovative TrIM Slices that implement the dataflow by a hardware perspective. In order to execute CLs, these slices are arranged in} 7 cores, each consisting of 24 slices of $3\times3$ PEs. \addtext{The cores compress psums spatially and the memory bandwidth is fully utilized by reading inputs once and broadcasting them to the different cores. Finally, different slices may cooperate with each other to manage large kernel sizes.}
    \item A reconfigurable shift-register buffer is presented. It handles different ifmap sizes for optimal SA utilization. This unlocks the TrIM-based architecture to cope with all the CLs of a generic CNN, being run-time reconfigured. \addtext{Additionally, it is worth pointing out that the use of shift-register buffers and the data broadcasting at the top level allow the architecture to manage the memory bottleneck at different hierarchical levels.}
    \item We perform a design space exploration for the VGG-16 CNN using a TrIM-based analytical model, where throughput, memory size and I/O bandwidth are considered.
    \item An FPGA-based implementation is presented to characterize the TrIM architecture in terms of resource occupation, performance, power and energy. \addtext{VGG-16 and AlexNet CNNs are considered, allowing validation on different ifmap and kernel sizes.} Comparisons with Eyeriss and state-of-the-art FPGA architectures are also presented, in order to highlight the benefits offered by the TrIM-based architecture.
\end{itemize}

The rest of the paper is organized as follows: in Section II state-of-the-art hardware implementation of SAs are reviewed; the TrIM-based hardware architecture is introduced in Section III; a design space analysis is presented in Section IV, considering throughput, I/O bandwidth and on-chip memory size; Section V reports the hardware implementation results, as well as comparisons with previous works; finally, conclusions are drawn in Section VI, with insights for future work.

\section{Related Works}
Given the way in which data flows between PEs, SAs are particularly suitable to manage Matrix Multiplications (MMs) between inputs and weights, being compatible with Fully-Connected Neural Networks (FCNNs). On the contrary, CNNs mainly execute convolutions, where inputs must be arranged in sliding windows before being processed with the respective weights. To achieve this, Conv-to-GeMM has been introduced to arrange sliding windows in a matrix form\cite{Chetlur_14}.

Several architectures based on SAs make use of Conv-to-GeMM\cite{Jouppi_17,Fornt_23,Ortega_24,Wu_24}. The Google Tensor Processing Unit (TPU)\cite{Jouppi_17} hosts a SA consisting of 256$\times$256 PEs. Based on the WS dataflow, inputs are supplied through First-In-First-Out (FIFO) buffers, and moved horizontally from left to right. The produced psums move vertically from top to bottom. In addition, an accumulation logic consisting of synchronization FIFOs and adders further process the psums, coming from the bottom side of the array, over multiple computational steps. According to this process, such a TPU is able to accommodate several NN workloads, including FCLs, CNNs, and Recurrent NNs. However, this flexibility has a cost in terms of memory capacity and accesses to cope with Conv-to-GeMM, thus degrading the energy efficiency. On the contrary, the accelerator proposed in\cite{Fornt_23} tackles the above issues by proposing an on-the-fly GeMM conversion. Specifically, a data feeder unit is interfaced to a SA of 16$\times$16 PEs and based on the OS dataflow. The data feeder includes an addressing logic to fetch inputs from proper regions of the main memory, as well as FIFOs to schedule inputs towards the array over time. In\cite{Ortega_24}, a Network-on-Chip based convolution is proposed to deal with GeMM. Inputs and weights are split into packets and provided to nodes consisting of parallel and distributed 3$\times$3 SAs. Despite this allows latency reduction, extra area is required to manage the NoC packets' scheduling. The work presented in\cite{Wu_24} deals with sparse CNNs, using a SA of 8 $\times$ 32 PEs. Ping-pong buffers assist local data reuse.

On another direction, some architectures\cite{Chen_17,Zhang_24,Sun_23,Liu_19} have proposed alternative ways to skip the disadvantages of GeMM conversion. Eyeriss\cite{Chen_17} is the first hardware chip proposing the RS dataflow, where both inputs and weights are circulated at PE level, while broadcasting between PEs at the SA level. An array consisting of 12$\times$14 PEs is interfaced to the global buffer through FIFOs for proper synchronization. Each PE, accommodating a two-stage multiplier and the adder, is equipped with scratch pads to circulate inputs and weights over different cycles. Moreover, a scratch pad for psums guarantees reuse over time. Each PE is also equipped with control logic to supervise the process. However, significant area is required to host the local scratch pads, other than negatively impacting the energy footprint due to the continuous switching activity of these memories. Moreover, the way in which the RS dataflow works requires that the area of the SA depends on the fmaps' sizes. As a result, to avoid a very large number of PEs, ifmaps' tiling must be carefully implemented. The RS dataflow is also exploited in\cite{Zhang_24}, where a 3-D 3$\times$3$\times$27 SAs is interfaced to a scheduling module to guarantee data reuse. thus avoiding inputs and weights circulation at the PE level. Differently from the RS dataflow, Sense\cite{Sun_23} offers a 32$\times$32 SA where different dataflows can be exploited: while weights can be retained at the PE level, psums can be either accumulated locally or forwarded vertically between PEs, thus also offering the possibility to meet the OS dataflow. However, the PE exhibits higher hardware complexity since it accommodates an extra buffer for psums as well as a sparsity management facility to skip zeroed computations from convolutions. The USCA architecture\cite{Liu_19} uses 64$\times$4 SAs, each having 4 PEs. Based on the OS dataflow, each PE is equipped with a selection logic to meet convolutions at different granularity, ranging from dense computations to dilated and transposed convolutions. PEs are equipped with sequential connections and skipping logic to properly place weights for sparse convolutions. However, other than requiring extra control logic, more connections must be placed at the boundaries of each PE, thus making routing challenging.

In the spectrum of hardware-based SAs, the TrIM-based architecture sits at the GeMM-free side.
\addtext{The TrIM dataflow combines IS and WS dataflows to maximize local data utilization, without affecting the external memory capacity. The TrIM dataflow operates on $K \times K$ PEs, interconnected with each other. Weights, inputs and psums go through the array as follows:}
\addtext{
\begin{itemize}
    \item Weights are grabbed from the memory and progressively stored at the PE level from \textit{top} to \textit{bottom}. The stored weights are thus kept stationary for the entire convolution.
    \item Inputs are initially read from the memory and moved into the array through \textit{vertical} links. Given that convolutional sliding windows share elements column-wise and row-wise, inputs are locally reused through \textit{horizontal} movements or \textit{diagonal} movements, respectively. Horizontal reuse is achieved from the rightmost PE to the leftmost PE, while diagonal reuse is supported by $K-1$ Shift Register Buffers. These buffers provisionally store row-wise shared inputs between sliding windows.
    These data are thus supplied diagonally to PEs. The composition of the three movements (i.e., vertical, horizontal, diagonal) configures a triangular shape, thus the name TrIM.
    \item Psums are accumulated from \textit{top} to \textit{bottom}. The psums generated at the bottom row of the array are finally reduced through an adder tree.
\end{itemize} 
}
\addtext{The local reuse of inputs make TrIM effective to mitigate the memory access count. For example, a $3 \times 3$ convolution over a $224 \times 224$ ifmap exhibits a negligible 1.8\% overhead.
In the following Section, we introduce a hardware architecture dealing with TrIM and suitable for the acceleration of CLs.}

\section{The TrIM Architecture} \label{TrIM_Architecture}
\begin{figure*}
\includegraphics[width=\textwidth]{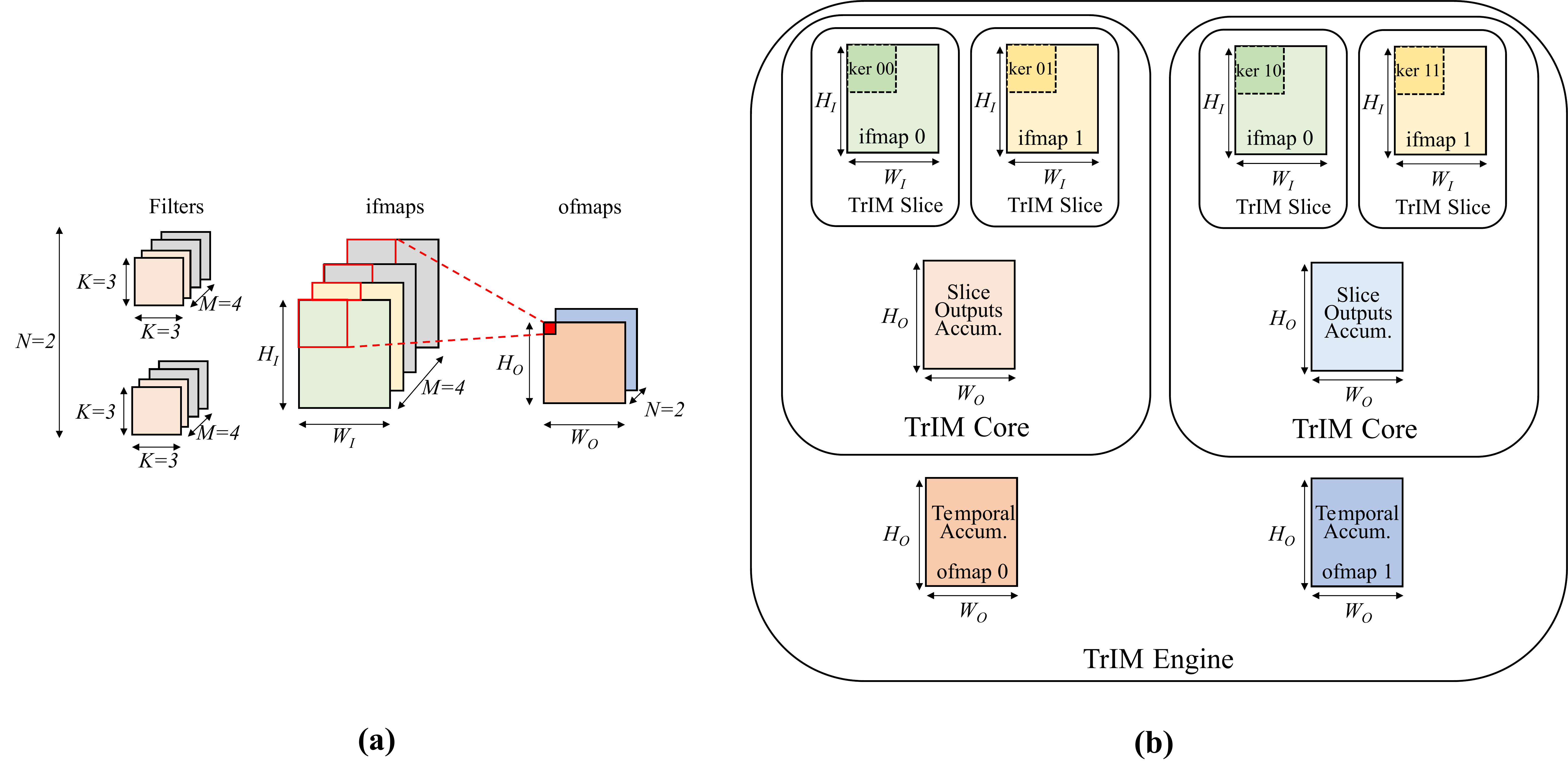}
\centering
\caption{Example of Convolutional Layer (CL) using the TrIM-based architecture: (a) The CL processes $M=4$ ifmaps and $N=2$ filters, each consisting of $M=2$ kernels. Each kernel accommodates $K \times K = 3 \times 3$ weights. As a result, $N=2$ ofmaps are generated; (b) example of processing of the first two ifmaps and the first two kernels of each filter. The TrIM Engine consists of two TrIM Cores, each having two TrIM Slices. The first TrIM Core processes ifmaps 0 and 1, as well as the first filter (kernels 00 and 01). The second TrIM Core processes ifmaps 0 and 1, and the second filter (kernels 10 and 11). Both TrIM Cores generate provisional ofmaps resulting as accumulations of the psums coming from their respective slices. Finally, the TrIM Engine is also responsible to accumulate over time the provisional ofmaps coming from the cores. In the specific case, the TrIM Engine accumulates the results from ifmaps 0 and 1, with the results from ifamps 2 and 3 in the subsequent iteration (not shown in figure).}
\label{TrIM_CL_Ex}
\end{figure*}
\begin{figure*}
\includegraphics[width=\textwidth]{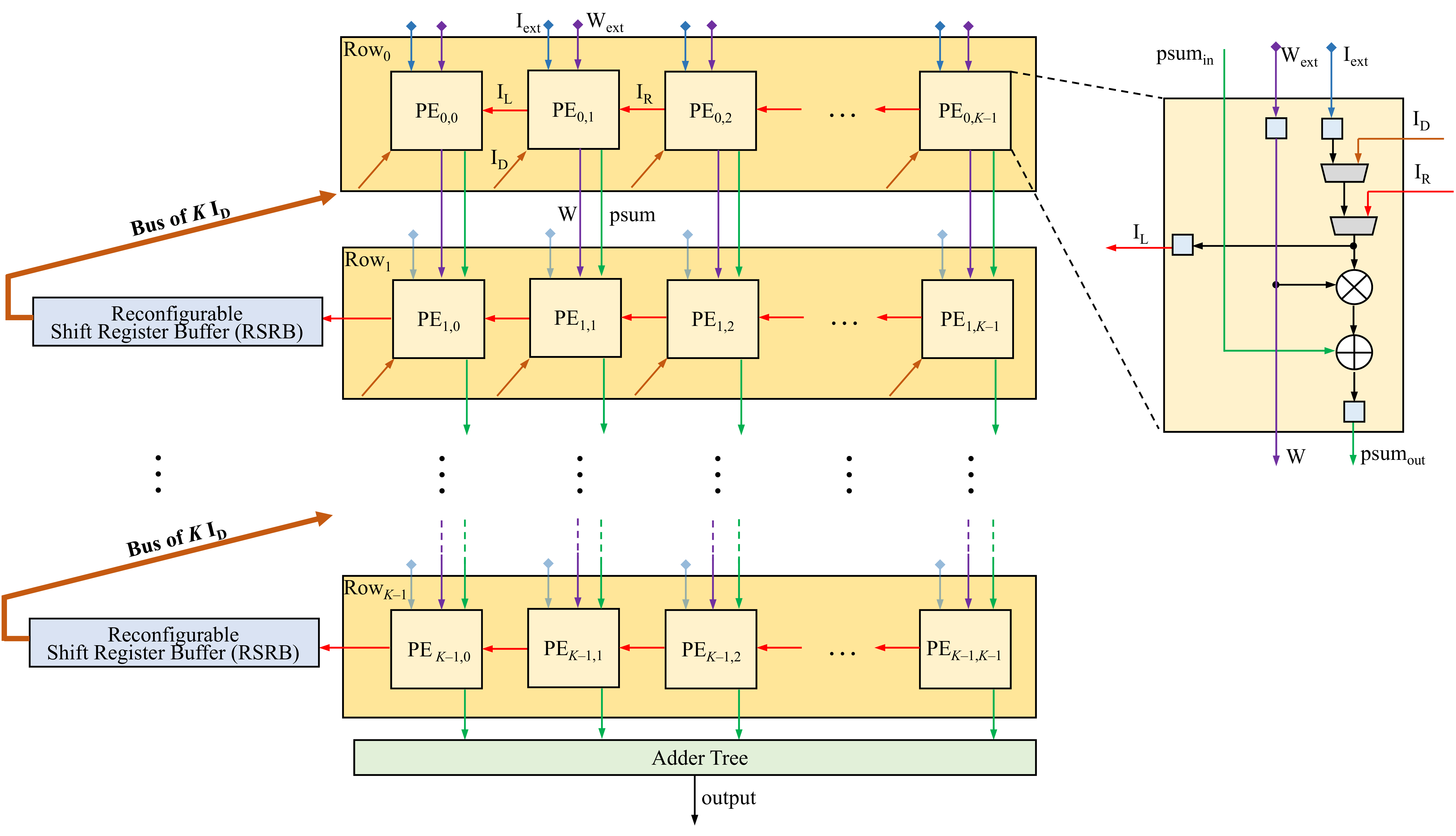}
\centering
\caption{The TrIM Slice. This consists of $K$ rows of $K$ Processing Elements (PEs) each, as well as $K-1$ Reconfigurable Shift Register Buffers (RSRBs) to facilitate the diagonal movement of inputs. For each PE, the arrows with different colors refer to a specific data/movement: the blue arrows indicate the external inputs supplied to the array; the purple arrows refers to the weights first supplied externally and, then, propagated from top to down; (c) the red arrows indicate the right-to-left movement of inputs; (d) the brown arrows indicate the diagonal movement of inputs from RSRBs to PEs; (e) the green arrows refer to the psums accumulation from top to down. A detail of the generic PE is also reported: it consists of four registers (in light-blue), two multiplexer to route the correct input, and the Multiply-Accumulation (MAC) unit. Finally, it is worth underlining that the outputs from the RSRBs can be seen as buses of $K$ inputs, which in turn feed as many PEs placed on the top.}
\label{TrIM_Slice}
\end{figure*}
\begin{figure*}
\includegraphics[width=\textwidth]{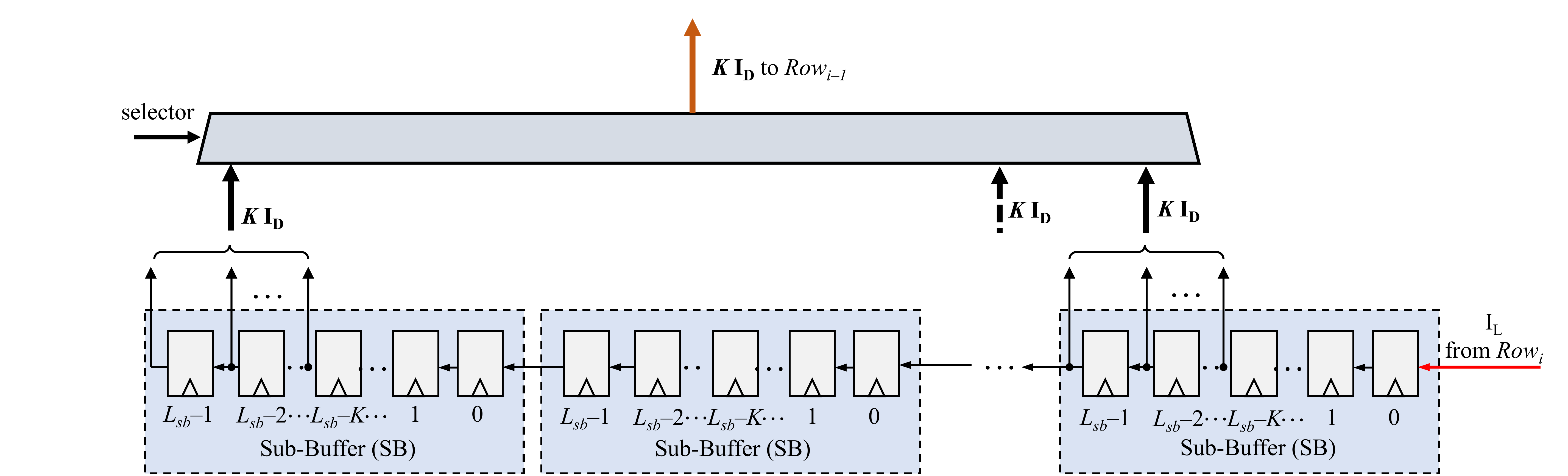}
\centering
\caption{The Reconfigurable Shift Register Buffer (RSRB). This consists of multiple Sub-Buffers (SBs), each having $L_{sb}$ shift registers, with $sb$ iterating over the number of SBs. $L_{sb}$ can be generic or customized parameters. Each SB provides the leftmost $K$ inputs to a multiplexer that, according to a given selection signal, provides the target set of $K$ inputs to the PEs placed in the top Row, in order to meet the diagonal movement of inputs.}
\label{SRB}
\end{figure*}
\begin{figure}
\includegraphics[width=\linewidth]{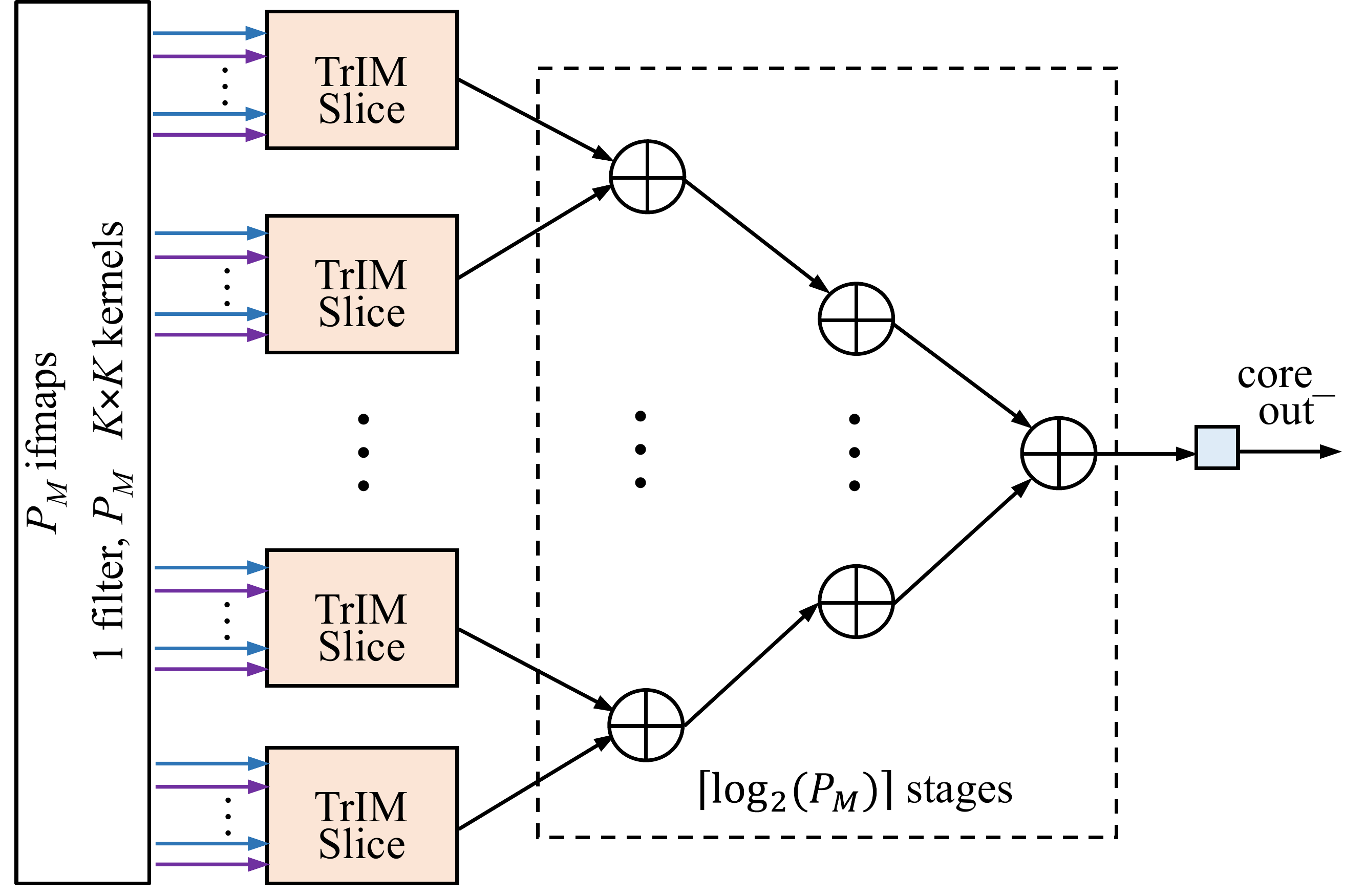}
\centering
\caption{The TrIM Core. It accommodates $P_M$ TrIM Slices, with $P_M$ being the number of parallel ifmaps processed by the accelerator. In addition, the TrIM Core spatially accumulates the outputs coming from the slices through an adder tree having a number of stages proportional to the logarithm in base 2 of $P_M$. The output is thus registered before being provided to further processing modules.}
\label{TrIM_Core}
\end{figure}
\begin{figure}
\includegraphics[width=\linewidth]{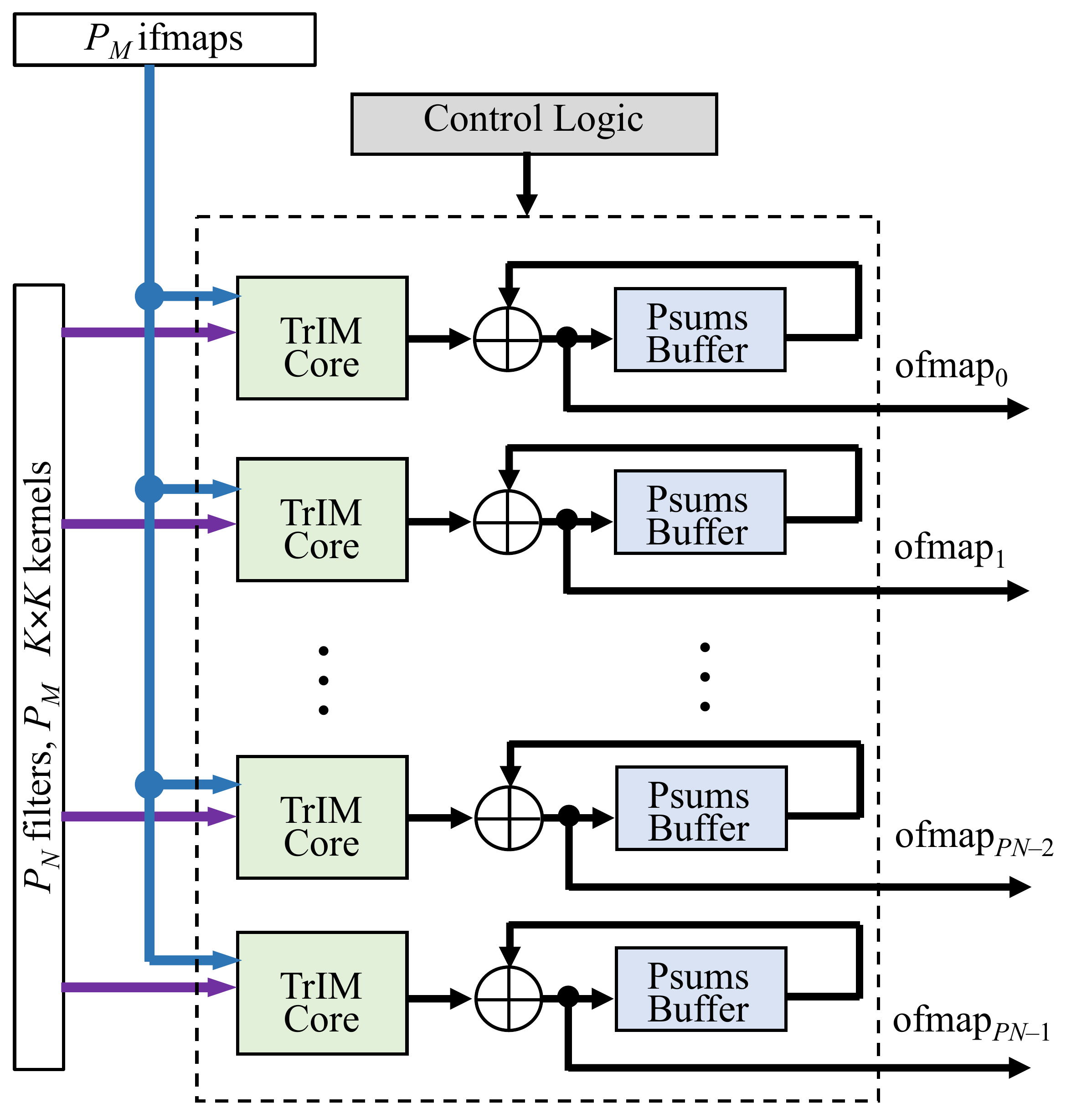}
\centering
\caption{The TrIM Engine. It accommodates $P_N$ TrIM Cores, with $P_N$ being the number of parallel filters (ofmaps) processed (generated) by the accelerator. In addition, an accumulation logic consisting of $P_N$ adders and psums buffers is provided. This allows the volumetric convolutions to be correctly executed over time. It is worth highlighting that the accumulation logic is required only when $P_N < N$. Finally, a control logic orchestrates the entire process.}
\label{TrIM_Engine}
\end{figure}
The TrIM-based AI hardware architecture is designed as a hierarchy consisting of three levels\addtext{, as reported in Fig. 2. The modules composing the hierarchy are the following:}
\begin{itemize}
    \item \textit{TrIM Slice}. This module is responsible for performing 2-D $K \times K$ convolutions. Weights are kept stationary by each Processing Element (PE), while inputs are reused between PEs by exploiting the innovative triangular movement. Specifically, each slice processes an independent ifmap and a specific kernel of one of the $N$ 3-D filters.
    \item \textit{TrIM Core}. This module accommodates multiple slices, each managing a specific ifmap and $P_M$ kernels belonging to one of the 3-D filters. $P_M$ is the number of kernels processed in parallel. Additional adders allow the provisional outputs from each slice to be accumulated, thus generating one ofmap.
    \item \textit{TrIM Engine}. This top-level module consists of $P_N$ cores, each dealing with an independent 3-D filter. $P_N$ is the number of filters processed in parallel and, in turn, the number of ofmaps generated in parallel. All cores use the same set of ifmaps.
\end{itemize}
Fig.~\ref{TrIM_CL_Ex} depicts an example of CL executed by the TrIM-based architecture. $M=4$ ifmaps and $N=2$ filters, each consisting of $M=2$ kernels, are considered. The TrIM Engine accommodates two TrIM Cores, each hosting two TrIM Slices. In the specific example, the first TrIM Core processes ifmaps 0 and 1, with kernels 00 and 01 (related to the first filter). The second TrIM Core processes ifmaps 0 and 1, with kernels 10 and 11 (related to the second filter). The TrIM Cores then produce provisional ofmaps, by summing up the psums coming from their respective slices. The TrIM Engine accumulates over time the provisional ofmaps coming from the Cores. 

In the following sub-sections, each part of the architecture is presented in detail.

\subsection{TrIM Slice}
The TrIM slice includes an array of $K \times K$ PEs, an adder tree and $K-1$ Reconfigurable Shift Register Buffers (RSRBs), as illustrated in Fig.~\ref{TrIM_Slice}. The computing array is organized in $K$ rows ($Row_i$, with $0 \leq i < K$), each having $K$ PEs. Preliminarily, the weights are taken from memory and provided to the PEs of $Row_0$ as groups of $K$ elements per cycle, then shifted from top to bottom. This process lasts $K$ cycles to ensure that the current kernel is completely stored at the slice level. Thus, inputs are first grabbed from the periphery ($I_{ext}$), moved in each row from right ($I_R$) to left ($I_L$), and forwarded to the closest SRB to complete the triangular movement. Each SRB dispatches $K$ inputs ($I_D$) to $Row_{i-1}$. Psums are first accumulated vertically through PEs and, finally, through an adder tree that finalizes the $K \times K$ convolution.

The generic PE is equipped with two input registers to store the external input $I_{ext}$ and the weight $W_{ext}$ (or $W$), respectively. Two cascaded multiplexers assist the PE in supplying the multiplier with the correct input ($I_{ext}$, $I_D$ or $I_R$). A subsequent adder accumulates the current product with the psum coming from the PE placed in the same column in $Row_{i-1}$. Finally, one output register delivers the current output either to the vertically-aligned PE in the $Row_{i+1}$ or to the Adder Tree, and another register provides the current input to the left PE. 

In order to make the slice agnostic to the ifmap size, the RSRBs is equipped with run-time reconfigurability. To achieve this, each RSRB consists of $W_{IM}$ registers, with $W_{IM}$ being the width of the largest ifmaps that typically constitutes the very first CL of a CNN. To manage the subsequent fmaps, having $W_I < W_{IM}$, the RSRB is split into multiple Sub-Buffers (SBs), each consisting of $L_{sb}$ registers (with $sb$ iterating over the SBs), as shown in Fig.~\ref{SRB}. While the majority of SBs simply forwards data from right to left, few SBs provide the data stored in the latest $K$ registers to a selection logic. This consists of a multiplexer, which provides a group of $K$ inputs to feed the $Row_{i-1}$, thus finalizing the triangular movement. It is worth underlining that the parameter $L_{sb}$ and the number of SBs feeding the multiplexer can be generic or customized.

The Adder Tree is supplied by $K$ psums coming from the bottom row of the array for the final accumulation. \addtext{Based on the binary-tree architecture,} it consists of $\lceil \log_2(K) \rceil$ stages, with the last one followed by an output register.

To conclude the description of the slice, some observations about the data representation and precision are needed. In the current version, the PEs support $B$-bit unsigned integer inputs and $B$-bit signed integer weights. As a result, signed integer psums are provided. In particular, the psums coming from the bottom row of the array have a bit-width equal to $2 \times B + K$. Then, after the adder tree, the current output is a signed integer consisting of $2 \times B + K + \lceil \log_2(K) \rceil$ bits.

\subsection{TrIM Core}
The TrIM core delivers 3-D convolutions by accommodating $P_M$ slices that operate in parallel, followed by an adder tree that further accumulates the $P_M$ provisional outputs, as depicted in Fig.~\ref{TrIM_Core}. With reference to a generic CL, the slices are fed by $P_M$ ifmaps (among the $M$ available), as well as by $P_M$ kernels from one of the $N$ filters. Taking into account that each slice operates on different ifmaps and kernels, the parameter $P_M$ is constrained by the I/O bandwidth that the architecture can sustain.  

The subsequent adder tree\addtext{, based on the binary-tree architecture,} produces a $2 \times B + K + \lceil \log_2(K) \rceil + \lceil \log_2(P_M) \rceil$-bit result (\textit{core\_out}), followed by an output register. Although the adder tree complexity can constraint the speed, pipelining can be added for shorter critical paths and better timing closure.

\subsection{TrIM Engine}
The TrIM engine is the top-level of the hierarchy as shown in Fig.~\ref{TrIM_Engine}, where multiple cores work in parallel on the same broadcast inputs. Specifically, $P_N$ cores are responsible for generating as many ofmaps, among the $N$ ofmaps to be produced. Considering that $P_N$ and $P_M$ are the parallelism parameters of the engine, up to $\lceil N/P_N \rceil \times \lceil M/P_M \rceil$ computational steps are required to finalize the computations. In the generic step, each core executes a 3-D convolution between one filter of $P_M$ $K \times K$ kernels and $P_M$ $H_I \times W_I$ fmaps. As a result, the $core\_out$ generated by the core is not final, but needs to be accumulated over other $\lceil M/P_M \rceil-1$ iterations. To achieve this, the engine associates an extra adder and a Psums Buffer to each core. This buffer needs to store up to $H_{OM} \times W_{OM}$ output activations, where  $H_{OM}$ and $W_{OM}$ represent the sizes of the largest ofmaps. Each activation is $2 \times B + K + \lceil \log_2(K) \rceil + \lceil \log_2(M) \rceil$-bit wide. For CLs handling smaller ofmaps, the buffer stores $H_O \times W_O$ output activations, where $H_O < H_{OM}$ and $W_O < W_{OM}$ are the sizes of the current ofmap. Meanwhile, the remaining $P_N-1$ cores processes different 3-D filters, but using the same set of $P_M$ ifmaps. In order to automatize the entire process over time, a Control Logic supervises the engine. Given that the scheduling of operations is the same for all the slices (thus, for all the cores), the cost of the controller is amortized by sharing it across the entire system. 

\section{Case Study: VGG-16} \label{case_study}
\begin{figure*}
\includegraphics[width=\textwidth]{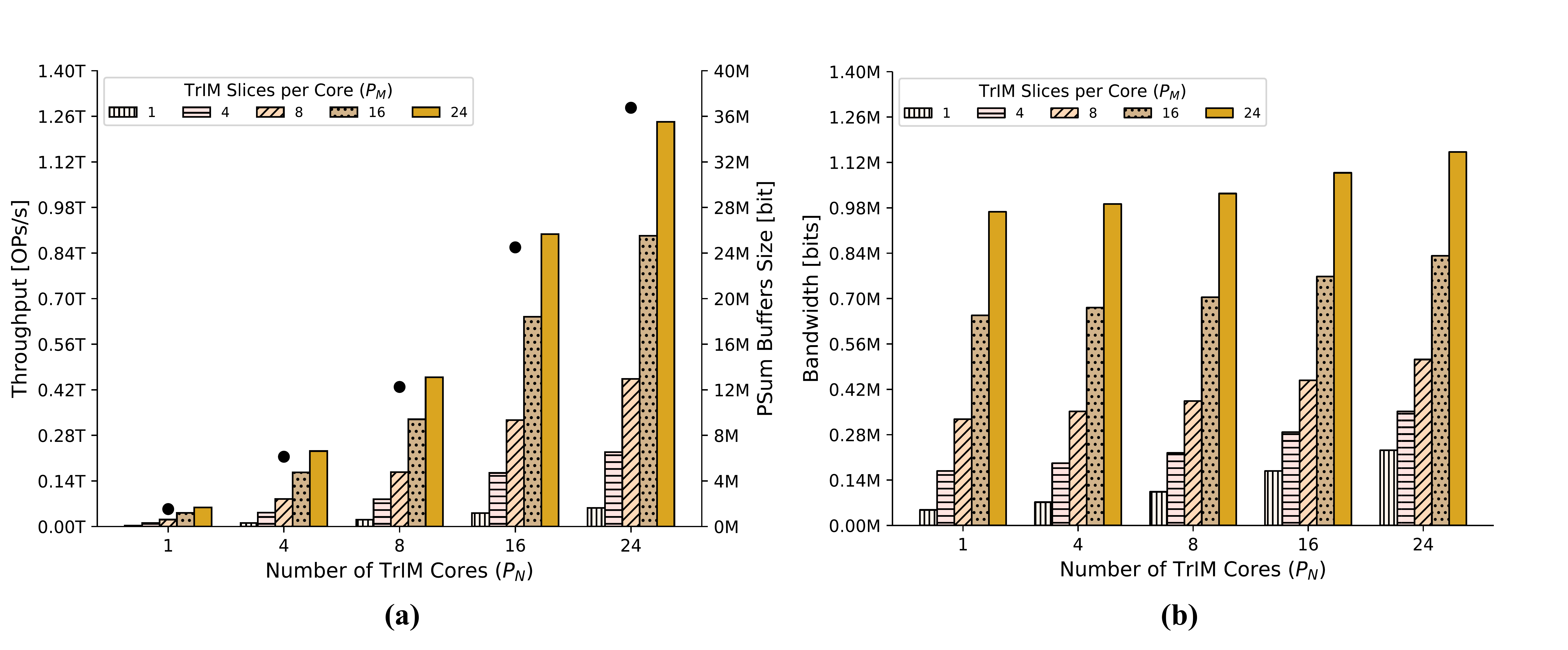}
\centering
\caption{Design Space Analysis: (a) Throughput and psums buffers size considering the parallelism factors $P_N$ and $P_M$. Five groups of bars refer to as many values of $P_N=[1,4,8,16,24]$ (i.e., parallel filters/ofmaps). For each group, five different bars are related to as many values of $P_M=[1,4,8,16,24]$ (i.e., parallel ifmaps), as reported into the legend. Throughput is expressed in terms of tera operations per second. The \addtext{points} indicate the psums buffers size, which is only dependent by $P_N$; (b) I/O bandwidth considering the parallelism factos $P_N$ and $P_M$, in Mbits. The bars follow the same format of the plot (a).}
\label{Parall_DS}
\end{figure*}
In order to showcase the capabilities offered by the TrIM-based architecture, we consider the Visual Geometry Group (VGG) CNN\cite{Simonyan_15} as a case-study. In particular, we refer to the VGG-16 architecture, where 13 CLs and 3 FCLs manage the image classification task. The focus of this research activity is oriented towards the hardware acceleration of the CLs only. 

We introduce some metrics to enable a design space exploration, in terms of achievable parallelism (i.e., the parameters $P_N$ and $P_M$). Three metrics are taken under consideration: throughput, psums buffers size, and I/O bandwidth.
In order to compute the throughput, the total number of operations and the execution time of the CNN under test are needed. Each CL performs 3-D convolutions between ifmaps and filters, demanding the following number of operations:
\begin{equation}
     OPs = 2 \times K \times K \times H_O \times W_O \times M \times N
\end{equation}
While this expression is general and independent by the architecture that processes the CLs' workload, the execution time is strictly constrained on the underlying architecture. As illustrated in Section~\ref{TrIM_Architecture}, the TrIM engine requires $\lceil N/P_N \rceil \times \lceil M/P_M \rceil$ computational steps to finalize each CL. Each step can be split into two phases: weights' loading and computation phase. During weights' loading, $P_N \times P_M \times K \times K$ $B$-bit weights must be provided to the architecture. Assuming that enough I/O bandwidth is available, it is supposed to fill one core with the needed weights every $K$ clock cycles. Therefore, it is expected to complete the entire phase in $P_N \times K$ clock cycles. Thus, the computation phase may start: after an initial latency dictated by the pipeline stages of the engine ($L_I$), $H_O \times W_O$ cycles are needed to finalize the convolutions. The number of clock cycles ($NC$) is given by expression~(\ref{clock_cycles}):
\begin{equation}
\label{clock_cycles}
     NC = L_I + \lceil N/P_N \rceil \times \lceil M/P_M \rceil \times (P_N \times K + H_O \times W_O)
\end{equation}
The execution time is retrieved by diving $NC$ by the clock frequency ($f_{CLK}$).

The psums buffers size is considered because on-chip memories contribute heavily to the energy footprint, other than being limited resources on FPGAs. Psums buffers store the provisional outputs from the TrIM cores, thus enabling temporal accumulations. Given the considerations reported in Section~\ref{TrIM_Architecture}, each psums buffer stores at most a $H_{OM}\times W_{OM}$ ofmap, where each activation is $2 \times B + K + \lceil \log_2(K) \rceil + \lceil \log_2(M) \rceil$-bit wide at least. Assuming 32-bit activations, enough to satisfy any on-chip accumulation, and taking into account that $P_N$ buffers are accommodated into the TrIM engine, the total psums buffer size is given by equation~(\ref{pbs_eq}):
\begin{equation}
\label{pbs_eq}
     Psums~Buffer~Size = P_N \times H_{OM} \times W_{OM} \times 32
\end{equation}

The I/O bandwidth is dictated by the transmission of weights, ifmaps and ofmaps. 
Weights are supplied before any processing: for each clock cycle, $P_M \times K \times B$ bits are moved. Conversely, the number of input activations vary over different cycles, to meet the triangular movement requirements at the slice level. Focusing on $K=3$, up to $P_M \times 5 \times B$ bits may be moved. Finally, the $P_N$ cores move as many $B$-bit quantized output activations, which can be transmitted to the periphery every $\lceil M/P_M \rceil$ steps. Considering that weights do not overlap with ifmaps/ofmaps transactions, the I/O bandwidth is constrained by the expression below:
\begin{equation}
\label{bw_eq}
     BW_{I/O} = (P_M \times 5 + P_N) \times B 
\end{equation}

Fig.~\ref{Parall_DS} shows the design space analysis, considering the metrics defined above. We span the analysis ranging the number of slices and cores from 1 to 24 and based on a clock frequency of 150 MHz, as for the FPGA implementation. As expected, the higher the number of cores ($P_N$) and the number of slices per core ($P_M$), the higher the throughput. The best-case with $P_N=P_M=24$ leads to a performance of 1243 Giga Operations per Second (GOPs/s). Furthermore, it is interesting to note how varying the number of cores and slices per core may bring to the same throughput. For example, an architecture consisting of 4 cores and 16 slices per core reaches the same throughput of an architecture managing 16 cores with 4 slices each. This because both the architectures use 576 PEs. However, the first solution with 4 cores is more efficient, since the psums buffer size is $4 \times$ lower, resulting in less area and energy requirements.

The I/O bandwidth increases significantly when varying the number of slices per core. This because the slices operate on different ifmaps, whereas multiple cores operate on the same set of ifmaps. 
Referring to the same example provided for the throughput, an architecture using 4 cores with 16 slices per core requires $2.3 \times$ more bandwidth than an architecture based on 16 cores with 4 slices per core. 

We can conclude that a TrIM-based architecture constrained on the psums buffer size can reach a competitive throughput when using more slices per core than cores. On the contrary, when the I/O bandwidth constraints the architecture, the same throughput may be achieved using more cores than slices per core.

\section{Hardware Implementation}
\begin{table*}[t]
  \centering
   \setlength{\tabcolsep}{4.5pt} 
   \renewcommand{\arraystretch}{0.9} 
  \caption{TrIM vs Eyeriss: Comparison on VGG-16}
  \begin{tabular}{c c c c c c c c c c c c c c c}
  \toprule
  \multicolumn{5}{c}{} & \multicolumn{5}{c}{TrIM} & \multicolumn{5}{c}{Eyeriss\cite{Chen_17}}\\
  \cmidrule(l){6-10} \cmidrule(l){11-15}
  \multicolumn{7}{c}{} & \multicolumn{3}{c}{\addtext{Memory Accesses [M]$^{\mathrm{a}}$}} & \multicolumn{2}{c}{} & \multicolumn{3}{c}{\addtext{Memory Accesses [M]$^{\mathrm{a}}$}} \\
  \cmidrule(l){8-10} \cmidrule(l){13-15}

   CL No. & $H_I \times W_I$ & \addtext{$K$} & $M$ & $N$ & GOPs/s & PE Util. & \addtext{On-Chip$^{\mathrm{b}}$} & \addtext{Off-Chip} & \addtext{Total} & GOPs/s$^{\mathrm{c}}$ & PE Util. & \addtext{On-Chip$^{\mathrm{b}}$} & \addtext{Off-Chip} & \addtext{Total} \\
\bottomrule
 \\ 1 & $224 \times 224$ & 3 & 3   & 64  & 51.8 & 0.13 & 0.00 & 13.57  & 13.57  & 13.7 & 0.93 & 43.81  & 7.70  & 51.51  \\
 \\ 2 & $224 \times 224$ & 3 & 64  & 64  & 368  & 1.00 & 0.57 & 102.79 & 103.36 & 13.7 & 0.93 & 477.14 & 27.00 & 504.14 \\
 \\ 3 & $112 \times 112$ & 3 & 64  & 128 & 387  & 1.00 & 0.27 & 49.96  & 50.23  & 13.7 & 0.93 & 271.44 & 16.70 & 288.14 \\
 \\ 4 & $112 \times 112$ & 3 & 128 & 128 & 387  & 1.00 & 0.68 & 95.33  & 96.01  & 13.7 & 0.93 & 495.48 & 24.25 & 519.73 \\
 \\ 5 & $56 \times 56$   & 3 & 128 & 256 & 396  & 1.00 & 0.33 & 48.51  & 48.84  & 27.2 & 0.93 & 145.57 & 10.10 & 155.67 \\
 \\ 6 & $56 \times 56$   & 3 & 256 & 256 & 432  & 1.00 & 0.66 & 94.71  & 95.38  & 27.2 & 0.93 & 259.22 & 16.10 & 275.32 \\
 \\ 7 & $56 \times 56$   & 3 & 256 & 256 & 432  & 1.00 & 0.66 & 94.71  & 95.38  & 27.2 & 0.93 & 255.46 & 15.40 & 270.86 \\
 \\ 8 & $28 \times 28$   & 3 & 256 & 512 & 422  & 1.00 & 0.33 & 52.44  & 52.77  & 52.8 & 1.00 & 89.08  & 8.90  & 97.98  \\
 \\ 9 & $28 \times 28$   & 3 & 512 & 512 & 422  & 1.00 & 0.70 & 103.72 & 104.42 & 52.8 & 1.00 & 157.88 & 14.30 & 172.18 \\
 \\ 10 & $28 \times 28$  & 3 & 512 & 512 & 422  & 1.00 & 0.70 & 103.72 & 104.42 & 52.8 & 1.00 & 141.23 & 11.40 & 152.63 \\
 \\ 11 & $14 \times 14$  & 3 & 512 & 512 & 389  & 1.00 & 0.17 & 33.05  & 33.23  & 57.4 & 1.00 & 32.69  & 3.15  & 35.84  \\
 \\ 12 & $14 \times 14$  & 3 & 512 & 512 & 389  & 1.00 & 0.17 & 33.05  & 33.23  & 57.2 & 1.00 & 29.68  & 2.85  & 32.53  \\
 \\ 13 & $14 \times 14$  & 3 & 512 & 512 & 389  & 1.00 & 0.17 & 33.05  & 33.23  & 57.2 & 1.00 & 28.95  & 2.80  & 31.75  \\
 \midrule
 \ Total &               &   &     &     & 391  & 0.93 & 5.44 & 858.63 & 864.06 & 24.5 & 0.94 & 2427.63 & 160.65 & 2588.28 \\
\bottomrule
    \multicolumn{14}{l}{\addtext{$^{\mathrm{a}}$Normalized to the same data width and a batch of 3 images.}} \\
    \multicolumn{14}{l}{\addtext{$^{\mathrm{b}}$Normalized to off-chip memory accesses.}} \\
    \multicolumn{14}{l}{$^{\mathrm{c}}$Retrieved from the number of VGG-16 operations per layer and following the given processing latency.} \\
  \end{tabular}
  \label{VGG16_TrIM_Eyeriss}
\end{table*}
\begin{table*}[t]
  \centering
   \setlength{\tabcolsep}{4.5pt} 
    \renewcommand{\arraystretch}{0.9} 
  \caption{TrIM vs Eyeriss: Comparison on AlexNet}
  \begin{tabular}{c c c c c c c c c c c c c c c}
  \toprule
  \multicolumn{5}{c}{} & \multicolumn{5}{c}{TrIM} & \multicolumn{5}{c}{Eyeriss\cite{Chen_17}}\\
  \cmidrule(l){6-10} \cmidrule(l){11-15}
  \multicolumn{7}{c}{} & \multicolumn{3}{c}{Memory Accesses [M]$^{\mathrm{a}}$} & \multicolumn{2}{c}{} & \multicolumn{3}{c}{Memory Accesses [M]$^{\mathrm{a}}$} \\
  \cmidrule(l){8-10} \cmidrule(l){13-15}

   CL No. & $H_I \times W_I$ & $K$ & $M$ & $N$ & GOPs/s & PE Util. & On-Chip$^{\mathrm{b}}$ & Off-Chip & Total & GOPs/s$^{\mathrm{c}}$ & PE Util. & On-Chip$^{\mathrm{b}}$ & Off-Chip & Total \\
\bottomrule
 \\ 1 & $227 \times 227$ & 11 & 3   & 96  & 2.13 & 1.00 & 0.08 & 8.44  & 8.52  & 51.1 & 0.92 & 17.92 & 2.50 & 20.42 \\
 \\ 2 & $27 \times 27$   & 5  & 48  & 256 & 179  & 0.57 & 0.21 & 3.50  & 3.71  & 45.7 & 0.80 & 28.64 & 2.00 & 30.64 \\
 \\ 3 & $13 \times 13$   & 3  & 256 & 384 & 390  & 1.00 & 0.11 & 14.85 & 14.95 & 54.9 & 0.93 & 15.09 & 1.50 & 16.59 \\
 \\ 4 & $13 \times 13$   & 3  & 192 & 384 & 402  & 1.00 & 0.07 & 11.20 & 11.27 & 56.1 & 0.93 & 10.44 & 1.05 & 11.49 \\
 \\ 5 & $13 \times 13$   & 3  & 192 & 256 & 399  & 1.00 & 0.05 & 7.52  & 7.57  & 59.8 & 0.93 & 5.36  & 0.65 & 6.01  \\
 \midrule
 \ Total &               &    &     &     & 12.9 & 0.91 & 0.53 & 45.50 & 46.03 & 51.5 & 0.88 & 77.45 & 7.70 & 85.15 \\
\bottomrule
    \multicolumn{14}{l}{\addtext{$^{\mathrm{a}}$Normalized to the same data width and a batch of 4 images.}} \\
    \multicolumn{14}{l}{\addtext{$^{\mathrm{b}}$Normalized to off-chip memory accesses.}} \\
    \multicolumn{14}{l}{$^{\mathrm{c}}$Retrieved from the number of AlexNet operations per layer and following the given processing latency.} \\
  \end{tabular}
  \label{AlexNet_TrIM_Eyeriss}
\end{table*}
\begin{table*}[t]
  \centering
  \renewcommand{\arraystretch}{0.9} 
  \caption{State-of-The-Art FPGA Architectures for Systolic Arrays}
  \begin{tabular}{c c c c c}
  \toprule
                                        & TVLSI '23\cite{Sun_23} & TCAS-I '24\cite{Wu_24} & TCAS-II '24\cite{Zhang_24} & This Work \\
    \midrule
     \\ Device                          & XCZU9EG   & XCZU3EG      & XCVX690T & XCZU7EV \\
     \\ Precision [bits]                & 16        & 8            & 16       & 8 \\
     \\ Processing Elements             & 1024      & 256          & 243      & 1512 \\
     \\ Systolic Array Dataflow         & OS,WS     & WS           & RS       & TrIM \\
    \midrule
     \\ LUTs                            & 348K                   & 40.78K & 107.17K & \addtext{194.35K} \\
     \\ FFs                             & N.A.$^{\mathrm{a}}$    & 45.25K & 34.45K  & \addtext{89.72K} \\
     \\ DSPs                            & 1061                   & 257    & 7        & 0 \\
     \\ BRAMs [Mb]$^{\mathrm{b}}$       & 8.82                   & 4.15   & N.A.     & 10.21 \\
    \midrule
     \\ Clock Frequency [MHz]           & 200                    & 150     & 150     & 150 \\
     \\ Peak Throughput [GOPs/s]        & 409.6                  & 76.8    & 72.9    & 453.6 \\
     \midrule
     \\ Power [W]                       & 11                     & 1.398   & 8.25    & \addtext{4.329} \\
     \\ Energy Efficiency [GOPs/s/W]    & 37.24                  & 54.94   & 8.84    & \addtext{104.78} \\
\bottomrule
  \multicolumn{5}{l}{$^{\mathrm{a}}$Not Available.} \\
  \multicolumn{5}{l}{$^{\mathrm{b}}$For AMD Xilinx FPGAs, BRAMs are usually expressed as number of 36Kb blocks. Here, the absolute size in Mb is reported.} \\
  \end{tabular}
  \label{SOTA}
\end{table*}
The TrIM-based architecture has been synthesized and implemented onto the FPGA of the AMD Zynq UltraScale+ XCZU7EV-2FFVC1156 MultiProcessor System-on-Chip (MPSoC). The design has been carried out through the AMD Vivado 2022.2 tool, and using Verilog as Hardware Description Language.

First, taking into account the design space reported in Section~\ref{case_study}, the optimal parameters $P_M$ and $P_N$ have been retrieved. $P_N$ is constrained by the on-chip memory. Considering that the XCZU7EV part offers 11 Mb of Block RAMs (BRAMs), using equation~(\ref{pbs_eq}), we got $P_N=7$, supposing that each psums buffer stores up to 224 $\times$ 224 output activations, which refer to the worst-case scenario (first two layers of VGG-16). 
The parameter $P_M$ strictly depends on the available I/O bandwidth. The XCZU7EV device is interfaced to a 64-bit DDR4 memory and exhibits a peak bandwidth of 19200 MB/s. \addtext{Supposing $f_{CLK}$ = 150 MHz, in line with state-of-the-art FPGA-based accelerators\cite{Sun_23,Wu_24,Zhang_24},} $BW_{I/O}$ = 1024, rounded to the closest power of 2. 
Therefore, considering equation~(\ref{bw_eq}) and $P_N=7$, we got $P_M=24$. Hence, a TrIM Engine hosting $P_N=7$ cores, each having $P_M=24$ slices, has been synthesized and implemented onto FPGA. \addtext{Placement and routing has been fully managed by Vivado. The architecture uses 194.35K Look-Up Tables (LUTs), 89.72K Flip-Flops (FFs) and 10.21 Mb of on-chip memory through Block RAMs (BRAMs). The pipeline depth of the TrIM Engine is 9 stages, of which: 5 stages for the TrIM Slices, 3 stages for the adder tree at the core level, and 1 stage for the temporal accumulation at the engine stage. The dynamic power is 4.329 W (86\% computations and data movement; 4\% on-chip memory; 10\% clock).}
The TrIM-based architecture achieves a peak throughput of 453.6 GOPs/s, since 1512 PEs are responsible for performing $7 \times 24$ parallel convolutions. 
With specific reference to the VGG-16 model, the engine guarantees a throughput of 391 GOPs/s, only 13.8\% lower than the peak throughput. This is motivated by high PE utilization, which reaches the 93\% on average. 

In order to highlight the advantages offered by TrIM in terms of performance and local data utilization, we present a comparative analysis with Eyeriss\cite{Chen_17}, state-of-the-art systolic array proposing the RS dataflow. \addtext{To showcase the effectiveness of TrIM in a wide spectrum of ifmap sizes and kernel sizes, the VGG-16\cite{Simonyan_15} and the AlexNet\cite{Krizhevsky_12} CNNs are benchmarked.} Table~\ref{VGG16_TrIM_Eyeriss} \addtext{and~\ref{AlexNet_TrIM_Eyeriss}} detail such a comparison, reporting (a) the specific configuration of each CL, (b) throughput, PE utilization and memory accesses. \addtext{Regarding the memory accesses, a breakdown between on-chip and off-chip memory is reported, and normalized to off-chip memory accesses.} 

\addtext{Regarding VGG-16, TrIM takes 78.6 ms (391 GOPs/s) to perform one inference step, while Eyeriss takes 1.25 s (24.5 GOPs/s). In terms of total memory accesses, TrIM requires $\sim3\times$ less than Eyeriss. This mitigation is motivated by the very limited contribution of on-chip memory. In fact, while Eyeriss requires a global buffer and scratch pads at the PE level, TrIM only uses a global buffer for psums. In detail, the $\sim94\%$ of equivalent on-chip memory accesses relates to scratch pads in the Eyeriss architecture. On the other hand, the use of a global buffer to store ifmap and weight tiles, along with data compression for ifmap sparsity, allows Eyeriss to save $5.3\times$ off-chip memory accesses than TrIM. However, this is not sufficient to hide the on-chip memory contribution, which is $\sim15\times$ higher.}

\addtext{To cope with the different kernel sizes required by AlexNet, the TrIM architecture splits large kernels in $3 \times 3$ tiles. For example, $P_M$ $5 \times 5$ kernels are split in 4 groups of $P_M$ tiles each. Each group is processed by a TrIM Core and the psums are accumulated at the top level to produce the ofmap associated to the entire $5 \times 5$ kernel. TrIM takes 103.1 ms to peform one inference step, while Eyeriss takes 26 ms. In terms of throughput, Eyeriss outperforms TrIM only in the first CL due to a mitigation of parallelism to split $11 \times 11$ kernels in $3 \times 3$ tiles. However, in the rest of layers ($5 \times 5$ and $3 \times 3$ kernels) TrIM outperforms Eyeriss up to $7 \times$. The drawback related to the first layer can be mitigated by more hardware specialization for the large kernels. 
In terms of memory accesses, TrIM confirms the trend achieved with the VGG-16 CNN. In the AlexNet case, TrIM uses $\sim1.8\times$ less memory accesses than Eyeriss. In fact, the equivalent on-chip memory accesses (scratch pads + global buffer) are $\sim10\times$ higher than off-chip accesses in Eyeriss. Also in this case, the motivation sits at the Eyeriss' PE level, where scratch pads dominate the on-chip accesses.}

The TrIM-based architecture is also compared to some state-of-the-art accelerators implemented onto FPGA \cite{Sun_23,Wu_24,Zhang_24}. Table~\ref{SOTA} summarizes the results, considering: (a) device type, data precision, number of processing elements and dataflow; (b) area occupation, in terms of LUTs, FFs, Digital Signal Processing slices (DSPs), and BRAMs; (c) clock frequency, peak throughput in GOPs/s; (d) power and energy efficiency.

Sense\cite{Sun_23} exploits a SA consisting of 32$\times$32 PEs. Considering that each PE must ensure both the WS and OS dataflow, other than guaranteeing the management of sparse convolutions, this results in higher hardware complexity and, in turn, more area. Indeed, even though TrIM makes use of $\sim$1.5$\times$ more PEs, Sense requires $\sim$79.1\% more LUTs. In addition, while TrIM completely manages multiply-accumulations through cheap logic offered by LUTs, Sense adopts DSPs. \addtext{TrIM, indeed, uses 8 bits for inputs and weights. Thus, computations do not require expensive 48-bit DSPs}. In turn, this affects the power consumption and, thus, the energy efficiency, being $\sim 3 \times$ lower than TrIM.

The architecture presented in\cite{Wu_24} proposes a SA using 8$\times$32 PEs and dealing with the WS dataflow. Among the counterparts, this hardware engine is the cheapest in terms of logic and on-chip memory, except for using 257 DSPs to implement the PEs. In turn, this results as the lowest power footprint. However, considering that \cite{Wu_24} reaches a peak throughput $5.9 \times$ lower than TrIM, this reflects in an energy efficiency $\sim 1.9 \times$ lower.

The SA introduced in\cite{Zhang_24} makes use of 3$\times$3$\times$27 PEs using the RS dataflow. In particular, it proposes a scheduling module that arranges the rows of data beforehand, thus simplifying the hardware complexity of PEs with respect to Eyeriss. However, despite using $\sim$6.2$\times$ less PEs than TrIM, this scales only $\sim$1.8$\times$ in terms of LUTs, because of extra logic requirements to manage sparse convolutions. Overall, TrIM outperforms \cite{Zhang_24} by $11.9 \times$ in terms of energy efficiency.

\section{Conclusion}
In this work, we present a hardware architecture to showcase the TrIM dataflow. The architecture is organized in a hierarchical manner, to effectively cope with convolutional layers. Reconfigurable shift register buffers are included to make the engine agnostic to different ifmap sizes. As a case study, the VGG-16 CNN is considered and an architecture of 1512 processing elements is implemented onto the XCZU7EV FPGA. The achieved peak throughput is 453.6 Giga Operations per Second. When compared to Eyeriss, a state-of-the-art accelerator based on the row stationary dataflow, the TrIM-based architecture guarantees $\sim 3 \times$ less memory accesses, confirming the effectiveness of the triangular input movement to maximize data utilization. In addition, the proposed architectures results up to $\sim 11.9 \times$ more energy-efficient than other FPGA counterparts.

Stimulated by these results, we plan the following future works:
\begin{itemize}
    \item Reduction of the shift register buffers' footprint through resource sharing. Considering that different processing elements may work on the same set of ifmaps, it is possible to share the same shift register buffers to assist the triangular movement of inputs. 
    \item Investigation about ifmap tiling to reduce the area required by the reconfigurable shift register buffers. At the moment, these local buffers are constrained on the largest ifmap size. By enabling tiling, it will be possible to reduce the number of registers per buffer, by further improving the energy efficiency of the architecture.
    \addtext{\item Insertion of a global buffer for ifmaps and weights, in order to reduce the count of off-chip memory access, thus strengthening the results with respect to previous art.}
    \addtext{\item In the direction of high energy efficiency, we are currently working on a tape-out of the TrIM-based architecture using a commercial CMOS technology, thus unlocking the applicability of TrIM in large-scale systems.}
\end{itemize}

\bibliographystyle{IEEEtran}
\bibliography{IEEEabrv,bib}

\begin{IEEEbiography}
[{\includegraphics[width=1in,height=1.25in,clip,keepaspectratio]{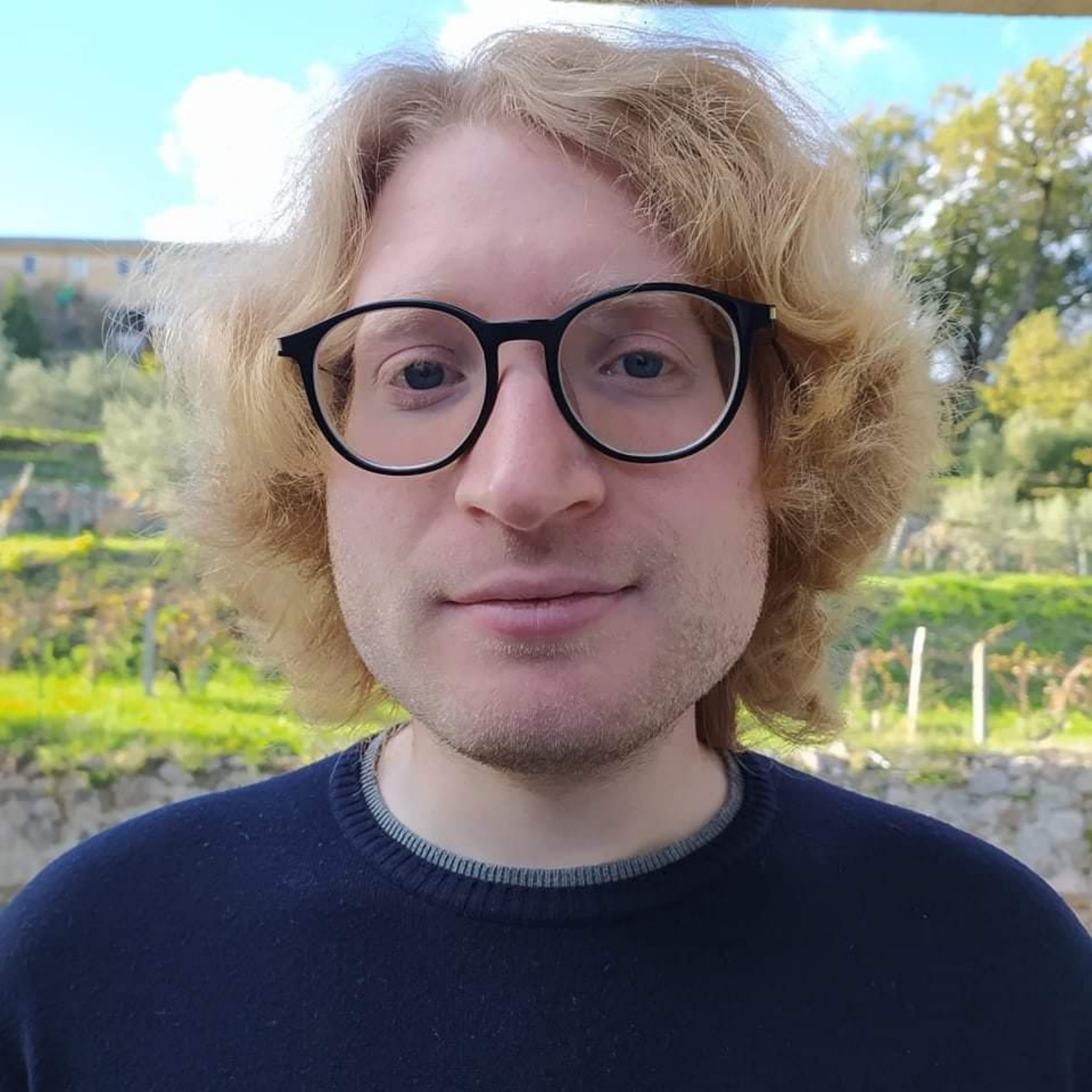}}]{Cristian Sestito}
(Member, IEEE) is a Research Fellow at the Centre for Electronics Frontiers CEF, The University of Edinburgh (UK). He received his BSc and MSc degree from University of Calabria (Italy), both in Electronic Engineering. He got his PhD in Information and Communication Technologies from the same university in 2023, focusing on Convolutional Neural Networks and their implementation on Field Programmable Gate Arrays (FPGA). In 2021/2022, Cristian was a Visiting Scholar at Heriot-Watt University, Edinburgh, working on neural networks’ compression. His research interests include digital design, embedded system design for AI on FPGA-based systems-on-chip, investigation of compression techniques (e.g., quantization), software simulators for neuromorphic AI.
\end{IEEEbiography}

\begin{IEEEbiography}
[{\includegraphics[width=1in,height=1.25in,clip,keepaspectratio]{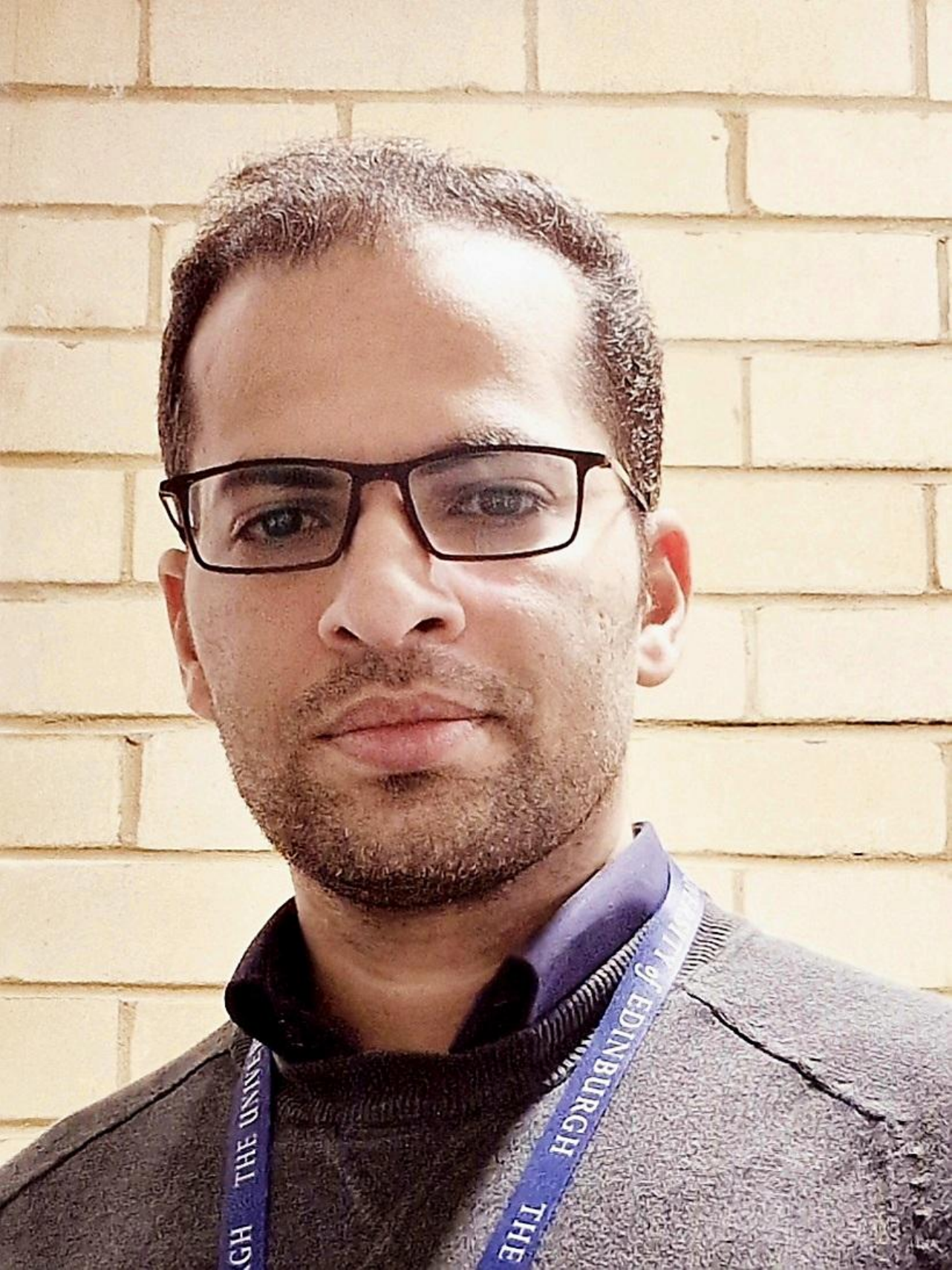}}]{Shady Agwa}
(Member, IEEE) is a Research Fellow at the Centre for Electronics Frontiers CEF, The University of Edinburgh (UK). He received his BSc and MSc degree from Assiut University (Egypt), both in Electrical Engineering. He got his PhD in Electronics Engineering from The American University in Cairo (Egypt) in 2018. Following his PhD, he joined the Computer Systems Laboratory at Cornell University (USA) as a Postdoctoral Associate for two years. In 2021, Shady joined the Centre for Electronics Frontiers at the University of Southampton (UK) as a Senior Research Fellow and then as a Research Fellow at the University of Edinburgh (UK). His research interests span across VLSI and Computer Architecture for AI using conventional and emerging technologies. His work focuses on ASIC-Driven AI Architectures with extensive expertise in In-Memory Computing, Stochastic Computing, Systolic Arrays, Beyond Von Neumann Architectures, Memories and Energy-Efficient Digital ASIC Design.
\end{IEEEbiography}

\begin{IEEEbiography}
[{\includegraphics[width=1in,height=1.25in,clip,keepaspectratio]{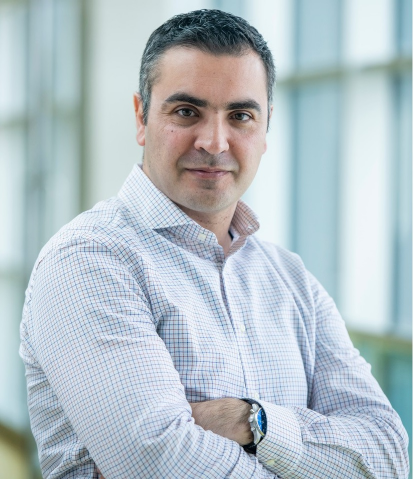}}]{Themis Prodromakis}
(Senior Member, IEEE) received the bachelor’s degree in electrical and electronic engineering from the University of Lincoln, U.K., the M.Sc. degree in microelectronics and telecommunications from the University of Liverpool, U.K., and the Ph.D. degree in electrical and electronic engineering from Imperial College London, U.K. He then held a Corrigan Fellowship in nanoscale technology and science with the Centre for Bio-Inspired Technology, Imperial College London, and a Lindemann Trust Visiting Fellowship with the Department of Electrical Engineering and Computer Sciences, University of California at Berkeley, USA. He was a Professor of nanotechnology at the University of Southampton, U.K. He holds the Regius Chair of Engineering at the University of Edinburgh and is Director of the Centre for Electronics Frontiers. He is currently a Royal Academy of Engineering Chair in emerging technologies and a Royal Society Industry Fellowship. His background is in electron devices and nanofabrication techniques. His current research interests include memristive technologies for advanced computing architectures and biomedical applications. He is a fellow of the Royal Society of Chemistry, the British Computer Society, the IET, and the Institute of Physics.
\end{IEEEbiography}

\end{document}